# A NOVEL MULTIPATH APPROACH TO SECURITY IN MOBILE AD HOC NETWORKS (MANETs)

Rangarajan A. Vasudevan
Department of Computer Science and Engineering,
Indian Institute of Technology, Madras
ranga@cs.iitm.ernet.in

Sugata Sanyal
School of Technology and Computer Science,
Tata Institute of Fundamental Research, Mumbai
sanyal@tifr.res.in

*Abstract-* **In this paper, we present a novel encryption-less algorithm to enhance security in transmission of data packets across mobile ad hoc networks. The paper hinges on the paradigm of multipath routing and exploits the properties of polynomials. The first step in the algorithm is to transform the data such that it is impossible to obtain any information without possessing the entire transformed data. The algorithm then uses an intuitively simple idea of a jigsaw puzzle to break the transformed data into multiple packets where these packets form the pieces of the puzzle. Then these packets are sent along disjoint paths to reach the receiver. A secure and efficient mechanism is provided to convey the information that is necessary for obtaining the original data at the receiver-end from its fragments in the packets, that is, for solving the jigsaw puzzle. The algorithm is designed to be secure so that no intermediate or unintended node can obtain the entire data. An authentication code is also used to ensure authenticity of every packet.**

## I. INTRODUCTION

Security of a network is arguably the most important issue in the world. Mobile Ad hoc NETworks (MANETs) are particularly prone to security attacks owing to their inherent nature of node mobility and the lack of a central governing infrastructure. The methods that have been employed for security in wired networks do not directly apply to ad hoc networks. Existing networks do not offer absolute security in the sense that they do not prevent the leak of information. A typical method that is used to prevent data from falling into wrong hands is encryption. Some encryption techniques like RSA involve algebraic operations with large numbers. The cost that has to be paid in implementing encryption in ad hoc networks is high owing to this computational complexity. So, encryption is not a viable solution for ad hoc network security. Another common feature of existing networks is the extensive use of single, shortest-path routing. Though this might offer the benefit of least data transmission time, the risk of compromise is higher compared to multipath routing.

This paper proposes a multipath approach to security in ad hoc networks. The essential idea is to break the data that is to be transferred into many packets. These packets when put together form the whole data only if done so in a particular way, just like in a jigsaw puzzle. The packets are sent using multipath routing. This reduces the chance of capture of all the packets by an adversary. We have incorporated efficient techniques that further enhance the security of the scheme, and at the same time implement the desired features. The tools that we have used in our algorithm are as follows: firstly, the concept of multipath routing; secondly, the All-or-Nothing Transform [3]; and thirdly, some useful properties of polynomials.

Multipath routing involves the transmission of data using more than one path from the sender to the receiver. This reduces the risk of an adversary monitoring all the traffic in all the paths originating from the sender. This is of course based on the assumption that an adversary cannot monitor all paths at the same time owing to the practical infeasibility. Multipath routing has attracted much attention including a paper on the use of multipath routing for secure data delivery [2]. However, our algorithm provides greater security than [2]. Moreover, our algorithm offers additional security features, and is designed for ad hoc networks.

The All-or-Nothing transform as described by Rivest in [3], describes a generic class of operations which when performed on raw data ensures that no information pertaining to raw data can be obtained from any part of the transformed data unless the entire transformed data is available. This couples well with multipath routing.

In this algorithm, polynomials have been used to design an efficient and secure way to provide information that is necessary to fit the pieces together to obtain the whole data. The properties of modulo division and efficient evaluation at fixed points are used. Shamir had used Lagrange interpolating polynomials in a secret-sharing system [1]. Krawczyk had proposed a system of distributing messages by an extension of Shamir's idea [7]. (The interested reader is referred to [6]).

This paper is organized as follows: in section II, we present a brief summary of research carried out in MANETs; in section III, we give the necessary foundation, and then describe the steps involved in our algorithm; section IV provides an analysis of and modifications/extensions to our algorithm; and, we conclude with section V.

## II. BRIEF REVIEW

A MANET is a collection of computers that communicate with each other through wireless channels, are not fixed to a location, and lack a central network governing body. Introductory and advanced material can be obtained from the sources [11], [12], [13], and [14].

There has been much research on security-related issues in MANETs [9], [15], and [16]. All the above consider only single-path routing. To the best of our knowledge, Lou et al [2] is the only paper that has a multipath approach to security, even though not for ad hoc networks in particular. They have extended the secret-sharing algorithm of Shamir and have used multipath routing. In essence, in a (T,N)-secret sharing system, the secure message is divided into N blocks of equal length such that from any T or more shares, it is possible to easily recover the message, while from any T-1 or less shares, it is computationally impossible to recover the message.

## III. THE ALGORITHM

### A. Background

1. The All-or-Nothing transform, as presented in [3], provides a preprocessing transformation f on the message M to be transferred. When applied to M, it ensures the following properties:

a. The transformation f is reversible; given the transformed message, one can obtain the original message;
b. Both the transformation f and its inverse are efficiently computable (that is, computable in polynomial time);
c. It is infeasible to compute any function of any message block if any one of the transformed message blocks is unknown.

An efficient implementation of a linear all-or-nothing transform is provided in [4]. However, to avoid known- and chosen- plaintext attacks, the input to this implementation has to be randomized. A point to be noted is that since there is no use of any secret key information that is used to transform the data, this does not qualify as encryption.

2. Given a polynomial f (), the problem of determining the roots of the polynomial is conditional upon the degree of the polynomial. A polynomial of degree less than 5 can be solved in the general case algebraically. However, by the use of Galois Theory, it can be shown that in the general case it is impossible to algebraically solve for the roots of polynomials of degree more than 5.

There are special cases though by which a polynomial of degree 5, say, can be solved. Suppose it is possible to guess a root of the polynomial owing to some interesting properties of the polynomial itself or the nature of the roots. Then, the problem can be reduced.

3. Authentication of a packet of data can be done using an algorithm like the HMAC-SHA1 [8]. A mechanism for message authentication, as outlined in [5], could be used. This algorithm takes as parameters the contents of the packet, the packet sequence number and a secret key known only to the sender and the receiver. At the sender-end, the Message Authentication Code (MAC) is calculated as indicated and is suffixed to the packet, and sent. At the receiver-end, using the secret key known to the receiver, the MAC is calculated and is compared with the MAC in the packet. If the two match, then the packet is considered to be from the sender only. Thus, false and/or modified packets can easily be detected by the receiver.

### B. Algorithm

Below, we present the algorithm in two parts - the sender's end and the receiver's end.

Sender's-end:
1: Apply All-or-Nothing transform to raw data D. Call the transformed data D'.

2: Split D' at N-1 arbitrary points resulting in data fragments $D'_1, D'_2, ..., $ and $D'_N$. Assign the values of the bit positions of the splits to $R_1, R_2, ..., R_{N-1}$.
3: Construct polynomial $P(x)$ using $R_i$'s as roots.
4: Evaluate $P(x)$ at the N different values $x_1, x_2, ..., x_N$. Denote the pairs as $(x_i, P(x_i))$.
5: Form N packets where for each packet, we do the following:
a. Assign a packet sequence number. Packets born out of the same split should be given same sequence numbers.
b. Fill the packet with $D'_i$ and the pair $(x_i, P(x_i))$
c. Calculate MAC of the packet with $K_0$ as secret key.
6: Send these N packets across N mutually disjoint paths for maximum security.

<u>Receiver's-end</u>:
1: On receiving the N packets, first check for matching sequence numbers. Then go to step 2.
2: For each packet, calculate MAC and verify authenticity. Then go to step 3.
3: Collect all the $(x_i, P(x_i))$ pairs from the packets and determine the polynomial $P(x)$ completely.
4: Calculate the roots of $P(x)$.
5: Using this knowledge of the positions of the splits, obtain the correct sequence of bits from the respective packets*. On concatenating them, and by the application of inverse-All-or-Nothing, the receiver can thus obtain the data.

The jigsaw puzzle is thus created at sender's end and solved at receiver's end.

* - The process of obtaining the correct sequence of bits from the packets is by trial-and-error. This is because there is no information available pertaining to the packet and the corresponding split that produced it. Unless the right combination of the packets is performed, the inverse-All-or-Nothing algorithm would return no definite pattern. Thus, the receiver can know whether he has combined the fragments in the right way. This issue is further discussed in the analysis section.

## IV. ANALYSIS

Firstly, the size of an ad hoc network is typically smaller than a wired network of today like the LANs in organizations. This combined with the mobility of the nodes make it difficult to find a large number of disjoint paths. For small values of N (lesser than 5,say), the computation involved as part of the algorithm in constructing the polynomial from its roots, evaluating the polynomial, interpolating the polynomial, and determining the roots of the polynomial are easily manageable. Also, the trial and error method for matching the split position to the right packet would be less complex.

However, as the values of N get larger, the computational cost increases. For values of N larger than 6, we are encountered with the problem of solving for the roots of a quintic (or higher degree) polynomial at the receiver's end. This is, however, not impossible because we know a crucial fact about the nature of its roots. The polynomial that is constructed at the sender's end has its roots bounded by 0 and (*size of packet – 1*). In the worst case, if an algebraic method cannot be used, we could do a traversal of all values in the range [0 - (*size of packet-1*)] and obtain all the roots. Thus, the worst case computational time is of O(*size of packet*). By the choice of the size of data and the availability of disjoint paths, we can arrive at an optimal value of N which gives the best tradeoff between computational complexity and security.

One more issue, as is clear, is that the complexity of this method is dependent on the packet size. Modern day wired networks are larger in size than a typical ad hoc network. This means that in wired networks, packet sizes are larger, and the number of disjoint paths is also larger. Thus, in scaling our algorithm to apply to large wired networks, the computational complexity becomes high.

Now, the security of the algorithm could be enhanced by the use of a prime number which should be a secret known only to the sender and the receiver. The polynomial is constructed as before but with an additional rule that all operations on the polynomial are performed modulo a large prime number *p*. (*p* needs to be greater than any of the coefficients of the polynomial and hence it needs to be large. For a more detailed explanation, refer [1] and [6].)

By implementing the above, in the unlikely case of an intruder gaining access to all the packets in all the paths, it is not feasible to determine the polynomial for lack of knowledge of *p*. On the other hand, in the case where this extension is not implemented, an intruder obtaining all the packets can use the computationally-feasible trial-and-error method to obtain the right combination of the fragments and thus get the entire data.

Also, using the above extension, the computational complexity on the part of the receiver to obtain the original data could be reduced. We could give provision for some additional bits in each packet where, these bits put together could give information that helps in exactly determining the order in which the fragments should be combined to get the whole. These extra bits, however, would not help the intruder since he has no knowledge of the prime *p*.

A possible method of providing the information is given as a modification to our earlier algorithm. Here, with each split, we send 2 pairs ($x_i$, $P(x_i)$) such that for a particular packet, $P(x)$ of that packet denotes a polynomial of degree 2 whose roots provide the starting and ending position of data in that packet. It is now possible to embed the data resulting by a split starting at any position in a packet while the remaining bits are filled randomly. The complexity involved in calculating the polynomial in the original algorithm can be avoided. However, now a quadratic polynomial should be calculated for every split that occurs. This method of embedding information provides additional security. The cost of implementing this extension to our algorithm is in terms of extra information exchanged and extra computations. If it is known that the topology and environment of an ad hoc network makes sure that it is absolutely impossible for an adversary to monitor all the paths originating from a sender, then this modification is unnecessary and can be avoided.

As can be noticed, the algorithm requires the a priori knowledge of a secret key $K_0$ for message authentication purposes. This key information could be exchanged either directly or by using an approach similar to our algorithm (using Shamir's Threshold scheme [1]).

Our algorithm is designed to work independently of the type of routing protocol employed. Hence, any multipath routing protocol like the Ad Hoc On-demand Multipath Distance Vector Routing [10] can be used.

A point to be noted in the algorithm is that the All-or-Nothing transform is used to essentially present an overall structure to the jigsaw puzzle. That is, it reduces the risk of the receiver putting together the pieces of the puzzle arbitrarily and still obtaining some data that was not intended. Also, any piece of the puzzle would not provide any clue to the puzzle itself. In this respect, our algorithm provides less information than a real-life jigsaw puzzle.

## V. CONCLUSION

Thus, we present a novel multipath approach to security in MANETs. The idea is to integrate multipath routing with the principle of a jigsaw puzzle. The data to be sent is broken arbitrarily into many packets, creating the set of pieces for a jigsaw puzzle. Sufficient information is provided securely so as to bring the pieces together in the right order at the receiver's end. Authentication codes are also added to all packets to provide more security. These packets are then sent to the receiver using multipath routing. Hence, without using encryption, we are able to achieve high levels of security with low computational complexity.

## ACKNOWLEDGEMENT

The authors would like to thank Tata Institute of Fundamental Research, Mumbai.